\documentclass[fleqn,10pt]{wlscirep}
\usepackage[utf8]{inputenc}
\usepackage[T1]{fontenc}
\usepackage{multirow}
\usepackage{color}

\newcommand{\beginsupplement}{
	\setcounter{table}{0}
	\renewcommand{\thetable}{S\arabic{table}}%
	\setcounter{figure}{0}
	\renewcommand{\thefigure}{S\arabic{figure}}%
}

\title{High-throughput Phenotyping of Nematode Cysts}

\author[1,+]{Long Chen}
\author[2,+]{Matthias Daub}
\author[3,+]{Hans-Georg Luigs}
\author[3]{Marcus Jansen}
\author[1]{Martin Strauch}
\author[1]{Dorit Merhof}
\affil[1]{Institute of Imaging \& Computer Vision, RWTH Aachen University, Aachen, 52074, Germany}
\affil[2]{Julius Kühn Institute: Federal Research Centre for Cultivated Plants, Elsdorf, 50189, Germany}
\affil[3]{LemnaTec GmbH, Nerscheider Weg 170, Aachen, 52076, Germany}

\affil[*]{corresponding.author@email.example}
\affil[+]{these authors contributed equally to this work}


\begin{abstract}
The beet cyst nematode (BCN) \emph{Heterodera schachtii} is a plant pest responsible for crop loss on a global scale. Here, we introduce a high-throughput system based on computer vision that allows quantifying BCN infestation and characterizing nematode cysts through phenotyping. After recording microscopic images of soil extracts in a standardized setting, an instance segmentation algorithm serves to detect nematode cysts in these samples. Going beyond fast and precise cyst counting, the image-based approach enables quantification of cyst density and phenotyping of morphological features of cysts under different conditions, providing the basis for high-throughput applications in agriculture and plant breeding research. 

In an evaluation using both manual cyst counts and manually annotated images, we show that the computer vision approach could compute accurate estimates of nematode cyst numbers, as well as accurate cyst segmentations. A phenotypical feature, cyst size, could be estimated as well for the automatic approach as for the manual ground truth, and it served to reveal differences between two nematode populations.
Source code and annotated evaluation data sets are freely available for scientific use.
\end{abstract}
\begin{document}

\flushbottom
\maketitle
%
%
\thispagestyle{empty}


\section*{Introduction}

\subsection*{Motivation}

Many nematode species, such as the beet cyst nematode (BCN) \emph{Heterodera schachtii}, are parasitic on plants 
and responsible for losses in crop yield that amount to annual financial losses of more than 150 billion USD~\cite{Singh2015}. Screening of soil samples for nematode infestation is a preventive measure and an integral part of pest management strategies in agriculture. It is also routinely performed by state institutions for import-export inspections of plants.

Cyst nematodes persist in the soil over many years as eggs inside a cyst, which is a protective shell formed by the remains of the former female body. Once a host plant germinates, juveniles are stimulated to hatch from the eggs and to leave the cyst. Juveniles move through the soil matrix, penetrate the roots and induce a so-called syncytium in the root tissue, causing damage to the plant. A juvenile that develops into a female produces offspring of up to several hundred eggs. After the eggs are developed, the female dies and the body wall turns into a brown, sclerotized cyst. 

The ability to quantify nematode infestation is a prerequisite for the evaluation of control measures targeted at nematodes,
as well as for the development of nematode-resistant plant breeding lines. For example, nematode population density may be estimated by cyst counting: The cysts need to be hand-picked from soil samples that still contain organic debris, where the amount of debris depends on the particular sample extraction method~\cite{Hallmann2020}. Manual counting
is a time-consuming task, and counting accuracy is affected by subjective decisions and the experience of the human counter in separating cysts from other particles of similar appearance, as well as by the ability to keep up concentration. Hence, only automated counting will be suitable for high-throughput applications in plant breeding.

Going beyond counting, also the extraction of phenotypical features related e.g.\ to size, shape and colour of the cysts is of interest. 
It could be shown that increased cyst size is an indicator of adaptation of a potato cyst nematode population to plant resistance~\cite{Fournet2016}, 
suggesting that the cyst phenotype can be informative for resistance applications. The phenotype can be determined from image data better than from other, e.g. genetic, modalities commonly used for nematode screening~\cite{Bogale2020}.

\begin{figure}[h]
    \includegraphics[width=\textwidth]{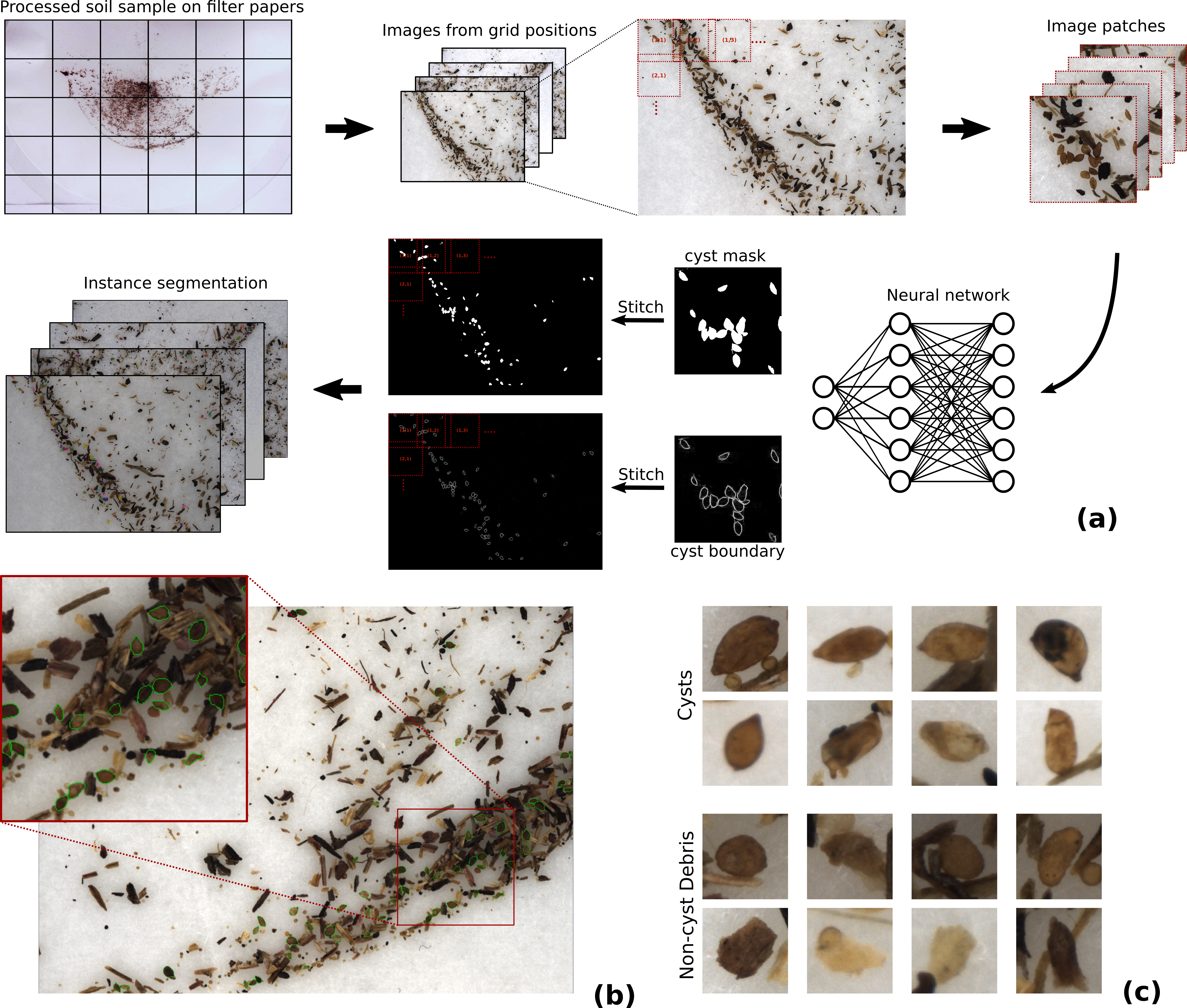}
   \caption{Introductory figure: \textbf{a)} Image processing pipeline: a grid of images covering the whole soil sample are processed separately. For each image, patches are processed by the neural network to generate cyst and boundary mask, from which the whole-image masks are stitched and further processed to get the final instance segmentation.  \textbf{b)} Example images containing nematode cysts (annotated in green) and organic debris particles. \textbf{c)} Examples of nematode cysts and debris particles that resemble the cysts.}
	\label{fig:introduction}
\end{figure}

Here, we introduce an automated system based on computer vision that serves as the basis
for extracting quantitative measures of nematode infestation from soil samples (Figure~\ref{fig:introduction}a). 
An optical microscopy readout followed by instance segmentation of the cysts amidst organic debris particles (Figure~\ref{fig:introduction}b) in only minimally processed soil samples enables fast processing in a high-throughput manner, while also providing access to phenotypical features. The system relies on a supervised learning model trained to detect \emph{Heterodera schachtii} nematodes, that are primarily parasitic on sugarbeets, but could be generalized to other nematode species with additional training data.

\subsection*{Contributions}

The main contributions of this paper are:
\begin{itemize}
\item A computer vision pipeline for instance segmentation of nematode cysts in microscopic images of extracts from soil samples (Methods, Figure~\ref{fig:introduction}a).
This is a challenging scenario with target instances immersed in a large number of distractors, i.e. soil particles, often with similar shape and colour (Figure~\ref{fig:introduction}b-c). 
\item An extensive evaluation showing that the system consisting of recording hardware and computer vision algorithms/software is able to detect nematode cysts with high accuracy in a high-throughput manner based on automatically recorded images of extracts from soil samples and without the need for laborious sample processing, such as hand-picking of cysts (Results).
\item Image data sets with manually counted cysts and manually annotated cysts masks.
\item A use case for cyst phenotyping, demonstrating how cyst populations from different soil types can be characterised morphologically (Results). 
\end{itemize}

\subsection*{Related work}

We have previously performed nematode cyst instance segmentation on images that are comparable to the image data from this paper, being recorded with a preliminary version of our image recording hardware: Chen et al.~\cite{Chen2019} have proposed a method for cyst instance segmentation that relies on instance proposals followed by instance classification with a SVM. The method was designed for small training data sets, where deep learning networks cannot play out their strengths. For the large annotated data sets from this paper, we employ deep learning strategies instead.

In this work, we focus on the detection of intact nematode cysts as they occur in unprocessed soil sample extracts. Once the cysts are broken up, e.g.\ through crushing, the eggs or juvenile nematodes contained therein will be released. Nematode egg detection has been performed by Akintayo et al.\cite{Akintayo2018} and Kalwa et al.~\cite{Kalwa2019}, and Chen et al.~\cite{Chen2020} have considered juvenile detection. We plan to include these additional instance segmentation scenarios into a future version of the high-throughput phenotyping system.

The nematode egg detection scenario from Akintayo et al.~\cite{Akintayo2018} is similar to our work in that relatively rare target objects need to be detected amidst a large number of cluttered distractor objects. The authors also employ a deep learning strategy for this detection task. While they consider processed samples to which fuchsin acid staining has been applied and that contain crushed cysts, our approach enables high-throughput applications without the need for physical treatment or for staining that can interfere with the phenotypical features. 

\section*{Results and Discussion}

In order to validate the cyst detection and segmentation performance of the automated system, i.e.\ image recording followed by the computer vision pipeline (Methods), we created two image data sets and associated evaluation scenarios:

\begin{itemize}
\item \emph{Cyst\_count} (Table~\ref{tab:data}): Cyst counting scenario. Images annotated with cyst numbers obtained by manual/visual counting.
\item \emph{Cyst\_segmentation}: Segmentation scenario. Images with manually annotated masks delineating the cyst boundaries. The data set consists of 229 images with
a total of $6331$ annotated cysts. 
\end{itemize}

\subsection*{\emph{Cyst\_count}}

In the \emph{Cyst\_count} scenario (Table~\ref{tab:data}), we compared manual and automatic cyst counts for a variety of soil sample types. Samples came from two different soil layers (top soil: 0-30 cm, sub soil: 31-60 cm) and had either high or low cyst density. For all combinations of these categories, we considered unprocessed samples with cysts and organic debris particles ("debris"), as well as clean samples where the cysts had been manually separated ("clean").

\begin{figure}[h]
   \includegraphics[width=\textwidth]{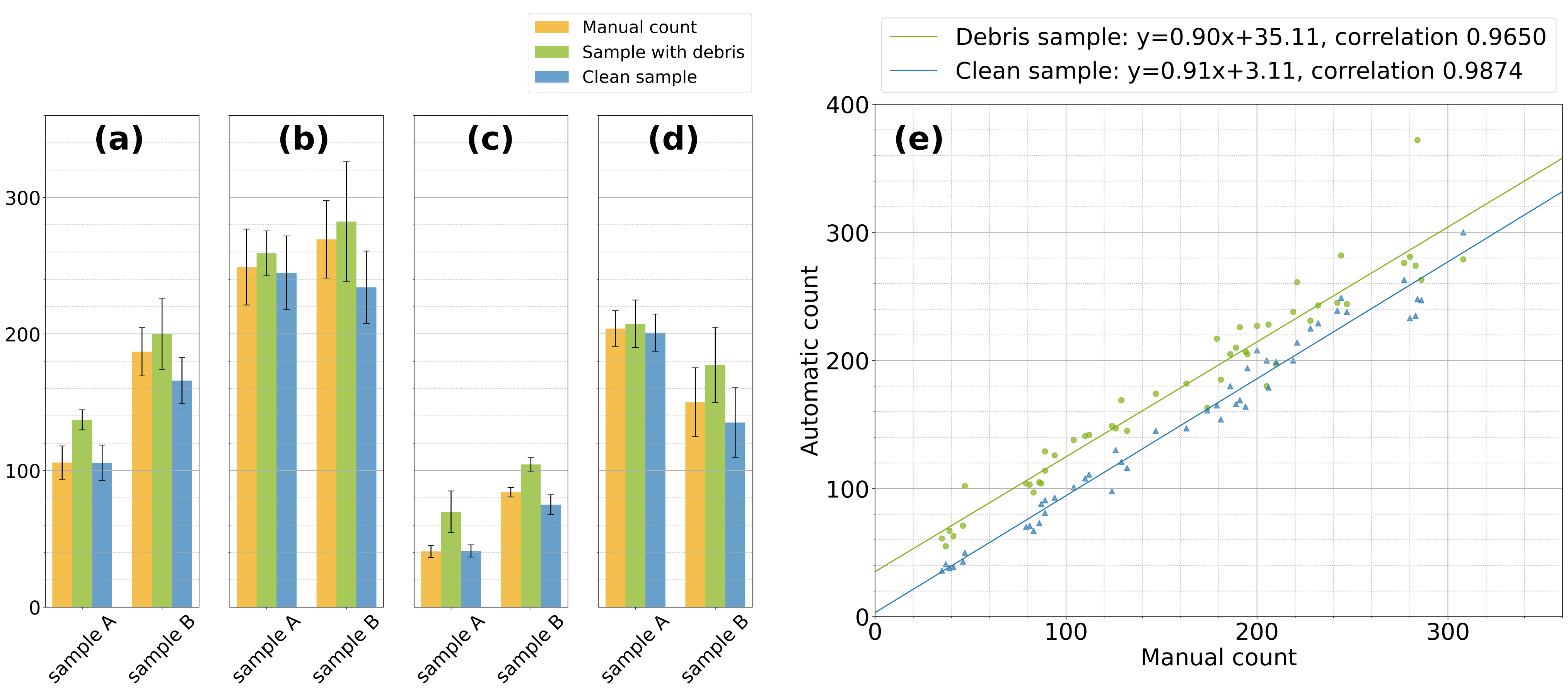}
   \caption{Automatic cyst counts are highly correlated with manual cyst counts. \textbf{a)} - \textbf{e)} Manual count, automatic count on a sample with debris and automatic count on a clean sample without debris for different types of soil samples: \textbf{a)} top soil, low cyst density (N = 2 locations * 6 replicates), \textbf{b)} top soil, high cyst density (N = 2*6), \textbf{c)} sub soil, low cyst density (N = 2*6), \textbf{d)} sub soil, high cyst density (N = 2*6). \textbf{e)} Manual count (N = 2*6*4) vs. automatic count on samples with debris and on clean samples. Pearson correlation coefficients for the correlation with manual count results are 0.965 and 0.987, respectively. 
   }
	\label{fig:cyst_count_manual}
\end{figure}

Across all soil sample types, manual cysts counts were similar to automatic cyst counts, both for the "debris" and for the "clean" case (Figure~\ref{fig:cyst_count_manual}a-d). Pearson correlation coefficients for the correlation between manual and automatic count were high, $0.965$ for the "debris" case and $0.987$ for the "clean" case (Figure~\ref{fig:cyst_count_manual}e), showing that the relative amount of cysts can be estimated reliably with the automatic method.

The number of cysts in a sample ranged between about 50 and about 300. Fitting a linear regression model to the manual (x) vs. automatic (y) count results revealed a small y offset of only $\approx +3$ for the automatic cyst count on the clean samples, while the y offset was $\approx +35$ for the debris samples (Figure~\ref{fig:cyst_count_manual}e), indicating highly accurate absolute count estimation on the clean samples, but also a consistent overestimation of cyst numbers in the debris samples, likely due to organic particles that resemble cysts.

%

%
%
%
%

\subsection*{\emph{Cyst\_segmentation}}

We next investigated how accurately the nematode cysts were segmented. For the \emph{Cyst\_segmentation} scenario, a total of 229 images of soil samples with between 3 and 94 cysts plus debris particles had to be segmented. A manually segmented ground truth was provided by a nematologist, while the automatic segmentation was performed with the computer vision pipeline that relies on the ResNet 101 as a backbone network (Methods). For comparison, we evaluated two further ResNets (50, 152) and a U-Net (Methods). 

\begin{figure}[h]
    \centering
	\includegraphics[width=\textwidth]{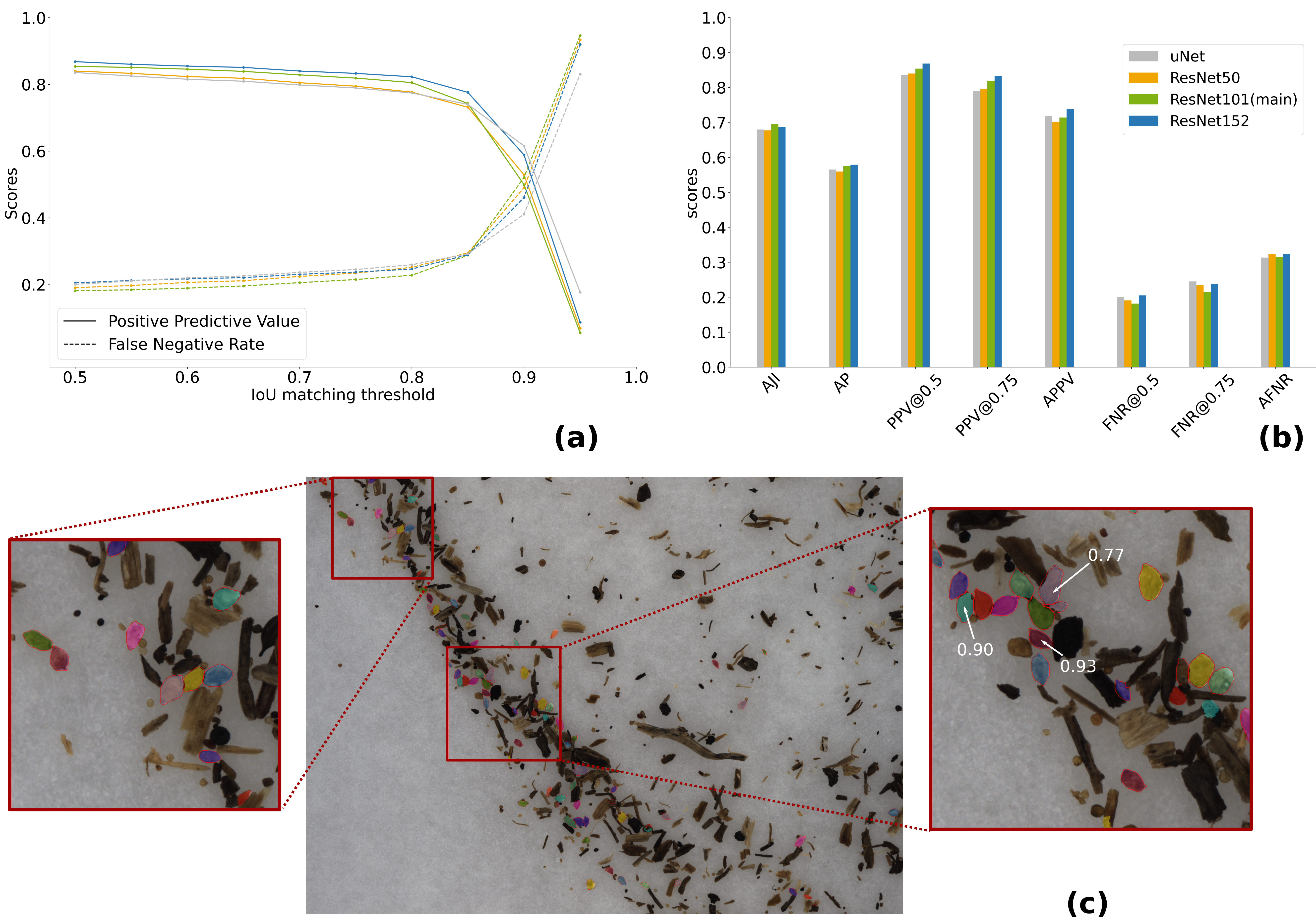}
   \caption{Cyst segmentation validation. \textbf{a)} PPR and FNR achieved by the computer vision pipeline (with ResNet 101 backbone) over a range of IoU thresholds. For comparison, we include three further network architecures (ResNet 50, 152, U-Net). \textbf{b)} Segmentation quality measures (Methods) for the computer vision pipeline with ResNet 101 and for the architectural variants. \textbf{c)} Qualitative examples for cyst segmentation masks obtained with the computer vision pipeline. Red lines indicate the ground truth boundary. Some instance masks are annotated with IoU scores.}
	\label{fig:segmentation}
\end{figure}

Due to the varying numbers of cysts per image, we chose an evaluation setting based
on cumulative scores, i.e.\ all cysts from all images were pooled. We measured segmentation accuracy by the Intersection over Union between the manually (M) and the automatically (A) segmented cyst mask, where $\text{IoU} : = (M\cap{}A) \, / \, (M\cup{}A)$. Employing a range of thresholds $\tau$ for the IoU, we counted a cyst as correctly segmented when $IoU \geq \tau$. We could thus compute the False Negative Rate (FNR, 1 - recall) at IoU thresholds $\tau = 0.5, \, \dots, 1.0$ for each of the four network architectures (Figure~\ref{fig:segmentation}a).

A similar trend could be observed for all four network architectures (Figure~\ref{fig:segmentation}a): 
For IoU thresholds up to $\tau=0.8$, the FNR, that measures the amount of missed cysts, remained constantly at about $0.2$, indicating that approximately 20\% of the cysts were missed and that $80\%$ could be segmented with such a good IoU. For higher $\tau$, the FNR increased, but was still about $0.5$ for $\tau=0.9$, 
i.e.\ half of the cysts could be segmented with a very high IoU of $0.9$. 

The Positive Predictive Value (PPV, precision), that measures the amount of all true positives among all true and false positives, exhibited the opposite trend, being constantly high at about $0.8$ up to $\tau=0.8$ and decreasing for higher $\tau$, crossing the FNR line at $\tau=0.5$. 

While the results for all network architectures followed the same trend, the ResNets had slightly, but consistently, lower FNRs for $\tau$ up to about $0.8$. However, for $\tau\geq0.9$, the U-Net had a slight advantage with respect to the FNR, indicating that the U-Net-based pipeline had failed to detect a few cysts, causing somewhat higher FNRs for the lower $\tau$, but in turn delivered a few more highly accurate segmentations with IoUs greater than $0.9$. 

In Figure~\ref{fig:segmentation}b, we report FNR and PPV for selected IoU thresholds, as well as a number of aggregated metrics, such as Average Precicision (AP), that summarize segmentation accuracy across several IoU tresholds (Methods). For all of these evaluation metrics, the four architectural variants performed on a similarly high level. 

In summary, we observed only minor performance differences for the the alternative networks, indicating that network architecture is not a critical part in the computer vision pipeline, as along as a reasonable choice is made, and that the default backbone network ResNet 101 is already a suitable architecture. Overall, the computer vision pipeline with the default backbone network yielded accurate instance segmentations where about 80\% of the cysts were segmented with a good IoU of about $0.8$ or higher. For qualitative results, see the segmentation examples with annotated IoU scores in Figure~\ref{fig:segmentation}c. 


\subsection*{Phenotyping nematodes}

Finally, we applied the high-throughput-system in order to provide a use case for cyst phenotyping. Based on the cyst segmentation masks for data set \emph{Cyst\_segmentation}, we computed cyst size (area in pixels) as a phenotypical feature, both for the $1366$ automatically segmented masks and for the $1399$ manually annotated ground truth masks. 

\begin{figure}[h]
    \centering
    \includegraphics[width= 0.5\textwidth]{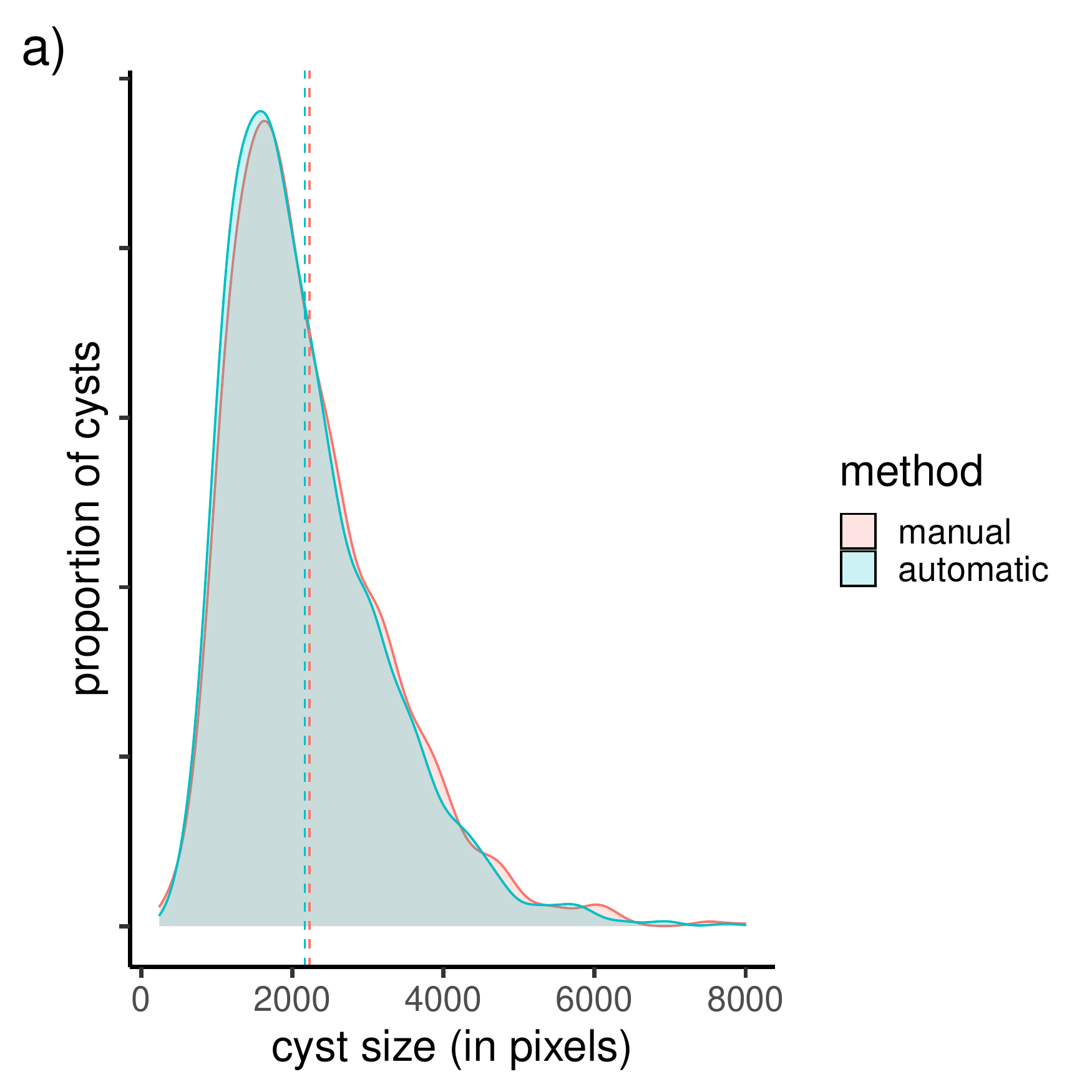}
    \includegraphics[width= 0.5\textwidth, page=2]{figs/cyst_size.pdf}
    \caption{Characterizing nematode cysts by cyst size (area in pixels) as a phenotypical feature.
    \textbf{a)} Cyst size estimates based on automatic segmentation are similar to those based on manually annotated cysts. On the \emph{Cyst\_segmentation} data (with debris), $1366$ (automatic) and $1399$ (manual) cysts, respectively, were detected and segmented. Density functions show the relative proportion of cysts of a certain size (x axis) for the respective distribution. The means of the distributions (automatic: $2164.86$, manual: $2226.71$) are marked by dashed lines.
    \textbf{b)} Characterizing two nematode populations in terms of cyst size. On the \emph{Cyst\_count} data (Table~\ref{tab:data}, clean samples without debris), $4503$ cysts were detected by the automatic method in top soil (0-30 cm) and $2714$ in sub soil (31-60 cm). Density functions visualize the relative proportion of cysts of a certain size. The means of the distributions (top soil: $2732.48$, sub soil: $2913.09$) are marked by dashed lines.
}
\label{fig:phenotyping}
\end{figure}

The density functions for the automatically and the manually determined size distributions reveal that, regardless of individual segmentation inaccuracies as measured by the IoU, the overall distribution of a phenotypical feature can be estimated reliably with the automatic high-throughput system (Figure~\ref{fig:phenotyping}a). Based on the Anderson-Darling k-sample test~\cite{Scholz1987} (R-package kSamples), the automatically and the manually determined cyst size distributions are not significantly different (test statistic T.AD=$0.6752$, p=$0.1731$). 

We next applied the high-throughput-system to the \emph{Cyst\_count} data (Table~\ref{tab:data}, clean samples) to analyze whether two nematode populations differ with respect to cyst size as a phenotypical feature. A total of $4503$ cysts could be detected and segmented in the top soil (0-30 cm) population, and $2714$ cysts in sub soil (31-60 cm). 

Indeed, the density distributions for cyst size as a feature revealed differences between the two populations (Figure~\ref{fig:phenotyping}b): The peak of both distributions was around 2500 pixels, but the shape of the distributions differed: In top soil, the relative amount of smaller cysts was larger, while the relative amount of larger cysts in the range of about 5000-10000 pixels was lower.\\ 
Based on the Anderson-Darling k-sample test~\cite{Scholz1987}, the null hypothesis that both populations, sub and top soil cysts, come from the same distribution, could be rejected (test statistic T.AD=$11.24$, p=$1.791e-05$). Hence, there is a significant difference between the two cyst populations, with more larger cysts in sub soil than in top soil. Also the average cyst size was larger in sub soil ($\approx 2913$) than in top soil ($\approx 2732$ pixels).

Further research is required to determine whether the observed differences generalize to soils from other locations. However, as a proof-of-principle, Figure~\ref{fig:phenotyping} demonstrates how automated cyst phenotyping can help to reveal even small, but consistent, morphological differences in cyst populations.
The high-throughput system enables processing of large numbers of samples, and could thus be employed for large-scale screening studies on soils from various geographic locations.

\section*{Conclusions}

We have introduced and evaluated a high-throughput phenotyping system (Methods) for nematodes in extracts from soil samples. At the core of the automated system lies a computer vision pipeline that achieves robust and accurate instance segmentation of nematode cysts in cluttered object collections with many remaining debris particles that often resemble the cysts in terms of shape and colour (Figure~\ref{fig:introduction}b-c). 

We have validated the automated system on a large number of images of soil sample extracts from different soils, comparing its results both against manual cyst counts and against manually annotated cyst segmentation masks (Results). Manual and automatic cyst counts were highly correlated (Pearson correlation coefficient $>0.96$), and about 80\% of the automatic cyst segmentations achieved IoU scores of $0.8$ or higher. The large number of debris particles did cause false positive cyst detections and hence a slight overestimation of the number of cysts. However, individual detection errors had no influence on a global phenotypical statistic: The distribution of automatically computed cyst sizes was not significantly different from the cyst size distribution based on manually segmented cysts, 

Cyst size is a simple phenotypical feature that can be computed from segmentation masks, and we have used it as a proof-of-concept for automatic morphological characterisation of nematode cyst populations. Future work will be focused on developing phenotypical feature sets based on shape, colour and texture of the cysts. The high-throughput systems facilitates processing of large data sets from screening studies, enabling phenotypical characterisation of nematode cysts under different environmental conditions, in soils from different locations, or in a resistance situation in the face of partly nematode-resistant plants.

\newpage
\section*{Methods}

\subsection*{Sample preparation}

Soil extracts were prepared from samples taken in various cropped fields with a known infestation history of the beet cyst nematode \emph{Heterodera schachtii}. Samples came from different fields distributed in a traditional sugar beet growing area in the Rhineland (Western Germany)
and from an experimental field at the Julius Kühn-Institute field station in Elsdorf, also situated in the same growing region (50$^{\circ}$55’41.27’’N; 6$^{\circ}$33’75.27’’ E). 

Soil extracts for image acquisition were prepared using a sieve combination and a subsequent centrifugation flotation technique~\cite{Viaene2020}, employing a MgSO$_{4}$ solution ($1.26$ g/ml) to facilitate buoyancy of the organic fraction. For each sample, approximately $300$g of field fresh soil were processed, resulting in a soil extract containing several hundreds of cysts. After pouring the centrifugate into a funnel, cysts along with the remaining organic rest were collected on a white filter paper (185 mm, MN 616 Macherey Nagel, Germany). A second filter paper put underneath absorbed excess water to avoid reflectance effects on the sample surface caused by light arranged above the sample. 

\subsection*{Image recording}

\begin{figure}[h]
    \begin{center}
    \includegraphics[width=.5\textwidth]{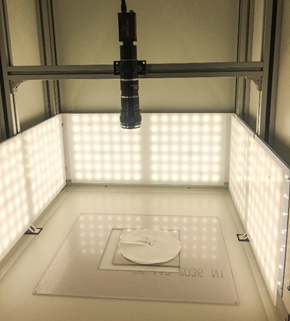}
    \caption{Microscopical image recording setup LemnaTec PhenoAIpert HM (High Magnification).}
	\label{fig:image_recording}
	\end{center}
\end{figure}

Images of the samples were recorded with a PhenoAIpert HM prototype (Figure~\ref{fig:image_recording}) by the company LemnaTec GmbH (Aachen, Germany). It consists of an industrial camera ($12.4$ megapixels) combined with a high-magnification lens system (1x magnification with 0.25x lower lens). The system has a field of view of $35$ mm $\times$ $25$ mm, pixel width being $0.009$ mm.

Samples were illuminated with horizontally oriented LEDs ($4000$ K) arranged around the sample stage. The sample stage consists of a sample holder that can be moved manually to a series of positions, such that the complete surface of the sample can be imaged in a series of photographs. All components are mounted inside an opaque cabinet that shields external light. 

\subsection*{Computer vision pipeline for cyst segmentation}

We employed a deep learning model with a convolutional neural network (CNN) in a semantic segmentation setting, performing pixel-level classification
of "cyst vs. noncyst", as well as "boundary vs. non-background".\\ 
As the images enclose a large field of view, they had to be processed patchwise in a split\&stitch manner due to GPU memory limitations. A semantic segmentation map is simpler and more efficient to stitch than the instance label map generated by detection-based approaches such as Mask-RCNN~\cite{mrcnn} and YOLO~\cite{yolov4}.\\ 
After stitching together the patch-level maps to obtain the image-level semantic segmentation maps for "cyst" and "boundary", we combined them to compute the final image-level instance segmentation map. For a flowchart of the computer vision pipeline, see Figure~\ref{fig:introduction}a).

\bigskip

\noindent \textbf{\textit{Semantic segmentation: Network architecture.}} The network takes three-channel RGB images as input. Raw pixel intensities are linearly scaled, such that every image has zero mean and a standard deviation of one. 

As the backbone network, we chose a ResNet~\cite{resnet} (ResNet 101). For Figure~\ref{fig:segmentation}, we also experimented with two further ResNet variants (ResNet 50, ResNet 152) and a  U-Net~\cite{unet}. In our U-Net implementation, we follow the typical U-Net architecture, but each convolutional layer is preceded by a batch normalization layer before activation.\\ ResNet has been proposed for and trained in an image classification setting. Here, we designed a segmentation model based on the original ResNet architecture: First, the fully connected layer and the global average pooling layer are removed, making the model fully convolutional. Thnen, we construct the decoding path with feature maps from the last convolutional layer of the conv1, conv2, conv3, conv4 and conv5 block, which have $1/2$, $1/4$, $1/8$, $1/16$ and $1/32$ of the original image resolution, respectively. The decoding path starts from the feature map with the lowest resolution and aggregates features iteratively by upsampling the lower resolution feature map and concatenating it with the one from the next resolution level:
\begin{align*}
    f'_{i} &= \mathcal{U}_i(f^*_{i}), \quad i=1,2,3,4,5 \\
    f_{i} &= \mathcal{C}(f^*_{i},\;f'_{i+1}), \quad i=1,2,3,4 \\
    f_{out} &= \mathcal{U}_{out}(f_1)
\end{align*}
\noindent where $f^*_{i}$ is the feature map extracted from the i-\textit{th} conv block of ResNet. $\mathcal{U}_i$ and $\mathcal{U}_{out}$ upscale the feature map via a transposed convolution layer with a $2\times2$ kernel and a $2\times2$ stride. $\mathcal{C}$ denotes a channel-wise concatenation of feature maps. For each concatenation, we keep the feature maps, $f^*_{i}$ and $f'_{i+1}$, balanced with the same channel number. Therefore, the filter number of the transposed convolution layer ${U}_i$ is determined by the channel number of feature map $f^*_{i-1}$.

At the output end, we pass the feature map $f_{out}$ to two independent convolutional layers, which are responsible for cyst and boundary recognition, respectively:

\begin{align*}
    S_{cyst} = \sigma(\;conv_{cyst}(f_{out})\;), \quad S_{boundary} = \sigma(\;conv_{boundary}(f_{out})\;)
\end{align*}

Both convolutional layers, $conv_{cyst}$ and $conv_{boundary}$, use a single $1\times1$ convolutional kernel, and the outputs are activated with a sigmoid function $\sigma(x)=1/(1+\exp(-x))$.

\bigskip

\noindent \textbf{\textit{Model training.}} For model training, we split the original images into $512\times512$ patches. Patches with more than $500$ cyst pixels were considered as containing cysts and used for training. 


The model was trained with the standard binary cross-entropy loss $ce(p1, p2) = -p_2\log(p_1) - (1-p_2)\log(1-p_1)$, both for cyst prediction and for boundary prediction:

\begin{align*}
    Loss = CE(S_{cyst},\;\textbf{C}) + CE(S_{boundary},\;\textbf{B}),
\end{align*}

\noindent, where $CE$ computes the mean value of $ce$ applied element-wise to every pixel. $\textbf{C}$ and $\textbf{B}$ are the binary maps for "cyst" and "boundary", respectively. To obtain the boundary map $\textbf{B}$, we performed a morphological dilation and erosion on the label map, using a disk-shaped structuring element of radius $2$. The difference between the dilated and eroded label map was taken as the boundary training map.  

Models were trained for $150$ epochs, using the RMSprop optimizer~\cite{rms} with a gradient decay of $0.9$. The learning rate was $0.0001$ initially and decreased exponentially to $90\%$ of its previous value every $10000$ steps. During training, we saved the best model based on the validation metric AJI~\cite{aji}. Model building and training was performed with Tensorflow (version 2.6).

\bigskip

\noindent \textbf{\textit{Split\&stitch processing.}} We employed a split\&stich strategy in order to be able to process images encompassing a large field of view. 
After splitting an image into patches, we computed score maps for "cyst" and "boundary" for each patch using the trained CNN model. Then, score maps at the patch level
were stitched together to obtain the score map at the image level. 

\begin{enumerate}
    \item \emph{Splitting and patch processing}: An image was split into overlapping patches of size $512\times512$. Each patch was processed separately by the trained CNN model, obtaining a cyst score map and a boundary score map.
    \item \emph{Stitching}: Patch predictions were stitched back to the original positions in the image. In the region of overlap, we averaged over the cyst scores from different patches, whereas for the boundary score, we took the maximum. This was motivated by the fact that the boundary is the more vulnerable structure and that successful separation of adjacent cysts relies on a continuous boundary segment between them. 
   \end{enumerate}

\noindent \textbf{\textit{Instance segmentation.}} Finally, we obtained an instance segmentation from the image-level semantic segmentation. 
\begin{enumerate}
 \item \emph{Instance separation}: We employed a threshold of $0.5$ to distinguish cyst pixels from background pixels in the cyst map. In the same way, boundary pixels were obtained from the score map. By excluding cyst pixels that were also boundary pixels, cyst instances could be separated spatially. Instances on the binary, thresholded map were then uniquely labelled by connected components labelling (8-neighbor connection).    
    \item Post-processing: Cyst pixels not assigned to any cyst instance were merged to the closest one. Furthermore, noisy predictions were removed based on a safe minimal size ($500$ pixels). 
\end{enumerate}


\subsection*{Evaluation metrics}

For the fully annotated data set \emph{Cyst-Seg}, we matched each predicted object to the ground truth object with the largest intersection over union (IoU). Using a certain IoU threshold, a match can be considered as a success (threshold exceeded) or a fail. The successful matches define the True Positives (TP), predicted objects that have no successful match are False Positives (FP), and ground truth objects without successful matches are False Negatives (FN). By accumulating these values through images of the whole data set, a precision metric (P) is defined as:

\begin{align*}
P_{IoU} = \frac{TP_{IoU}}{TP_{IoU}+FP_{IoU}+FN_{IoU}}.
\end{align*}

By sweeping over a range of IoU thresholds, an average precision under different level of matching rigor can be computed as an overall metric, taking into account the effect of segmentation accuracy:

\begin{align*}
AP = \frac{1}{|IoU|} \sum_{\substack{IoU}} \frac{TP_{IoU}}{TP_{IoU}+FP_{IoU}+FN_{IoU}},
\end{align*}

\noindent, where the threshold values range from $0.5$ to $0.95$ with a step size of $0.05$ in this work. Furthermore, we computed the positive predictive value (PPV) to quantify detection accuracy, and the False Negative Rate (FNR) to measure how many objects were missed:

\begin{align*}
PPV_{IoU} = \frac{TP_{IoU}}{TP_{IoU}+FP_{IoU}}, \quad
FNR_{IoU} = \frac{FN_{IoU}}{TP_{IoU}+FN_{IoU}}.
\end{align*}

We also report the Average PPV (APPV) and the Average FNR (AFNR) over different IoU matching thresholds. As a further measure for segmentation quality, we used the Aggregated Jaccard Index (AJI)~\cite{aji}. 

\newpage
\bibliography{main}



\section*{Acknowledgements}

This work was funded by the Germany Ministry of Education and Research (grant number 031B0474C).

\section*{Author contributions statement}

MD, HGL, MJ, LC, MS and DM planned research. LC developed machine learning models.
MD collected soil samples, designed the sample preparation procedure and created the annotated evaluation data.
HGL and MJ developed the optical recording setup and recorded images.
LC and MS analysed data and wrote the manuscript with contributions from the co-authors. All authors reviewed the final manuscript.
 
\section*{Additional information}

\textbf{Competing interests} The image recording system used in this work, PhenoAIxpert HM, is a commercial product distributed by LemnaTec GmbH.

\noindent \textbf{Data and code availability} Source code and evaluation data sets are available online for non-commercial use: LINK


\section*{Figures and Tables}

\begin{table}[ht]
\centering
\begin{tabular}{cccccc}
\hline
Data set & Sample name & Sample weight & \# Replicates & \# Images & Description\\
\hline
\multirow{8}{*}{Cyst-Count} & Top-Low-A Debris/Clean & 300g & 6 & 6x(30+12) & \multirow{2}{*}{top soil, low cyst density}\\
& Top-Low-B Debris/Clean & 300g & 6 & 6x(30+12) & \\\cline{2-6}
& Top-High-A Debris/Clean & 300g & 6 & 6x(30+12) & \multirow{2}{*}{top soil, high cyst density}\\
& Top-High-B Debris/Clean & 300g & 6 & 6x(30+12) & \\\cline{2-6}
& Sub-Low-A Debris/Clean & 300g & 6 & 6x(30+12) & \multirow{2}{*}{sub soil, low cyst density}\\
& Sub-Low-B Debris/Clean & 300g & 6 & 6x(30+12) & \\\cline{2-6}
& Sub-High-A Debris/Clean & 300g & 6 & 6x(30+12) & \multirow{2}{*}{sub soil, high cyst density}\\
& Sub-High-B Debris/Clean & 300g & 6 & 6x(30+12) & \\
\hline
\end{tabular}
\caption{Samples used for cyst count validation. The data set contains top soil samples (0-30 cm) and sub soil samples (31-60 cm) of the sugar beet field with different cyst density. Sample A and B were collected after (in March) and before (in September) sugar beet planting period, respectively.}
\label{tab:data}
\end{table}


\beginsupplement



\end{document}